\documentclass[11pt]{article}
\usepackage[dvips]{graphicx}
\usepackage{a4wide}
\def\be{\begin{equation}}
\def\ee{\end{equation}}
\def\bea{\begin{eqnarray}}
\def\eea{\end{eqnarray}}
\def\a{\alpha}

\def\b#1{\beta_{#1}}

\def\f#1#2{\frac{#1}{#2}}

\title{QCD perturbation theory at large orders with large renormalization scales in the large $\b{0}$ limit.}
\author{K. Van Acoleyen and H. Verschelde\\Department of Mathematical Physics and Astronomy, \\
University of Ghent, \\Krijgslaan 281 (S9), 9000 Ghent,
Belgium.\\Email: \email{karel.vanacoleyen@rug.ac.be}, \email{
henri.verschelde@rug.ac.be}}
\begin{document}
\title{ \textbf{QCD perturbation theory at
large orders with large renormalization scales in the large
$\b{0}$ limit.}}
\author{K. Van Acoleyen and H. Verschelde\thanks{karel.vanacoleyen@ugent.be and henri.verschelde@ugent.be} \\
{\small {\textit{Ghent University }}}\\
{\small {\textit{Department of Mathematical Physics and Astronomy,
Krijgslaan 281-S9, }}}\\
{\small {\textit{B-9000 Gent, Belgium}}}}\maketitle \vspace{5cm}
\noindent {\bf Abstract.} We examine the QCD perturbation series
at large orders, for different values of the 'large $\b{0}$
renormalization scale'. It is found that if we let this scale grow
exponentially with the order, the divergent series can be turned
into an expansion that converges to the Borel integral, with a
certain cut off. In the case of the first IR renormalon at
$2/\beta_0$, corresponding to a dimension four operator in the
operator product expansion, this qualitatively improves the
perturbative predictions. Furthermore, our results allow us to
establish formulations of the principle of minimal sensitivity and
the fastest apparent convergence criterion that result in a
convergent expansion.

\makeatother
\newpage

\section{Introduction}
It has been known for many years that the QCD perturbation series
suffers from factorial divergencies \cite{th77}.  At the same
time, we know that the perturbation theory itself does not tell us
the whole story and one needs to consider non-perturbative
contributions. Because of the asymptotic freedom, we expect these
contributions to become more important for lower energies. This is
confirmed by the operator product expansion (OPE) \cite{w69}, the
general framework that is used to parameterize the
non-perturbative contributions in several power corrections,
corresponding to several condensates \cite{svz79}. In fact, one
can show that for one source of divergencies in the perturbation
series, the so called IR renormalons, there exists a one to one
correspondence with these condensates \cite{m85,b98}. This is
traced back to the Borel insummability of the perturbation series,
leaving ambiguities in the definition of its Borel sum, that can
be exactly compensated by different values of the
(non-perturbative) OPE condensates. Another crucial problem with
the Borel sum, is that the Borel integral is expected to diverge
at infinity \cite{th77}. However, for now we will disregard this
problem and assume that the integral converges at infinity, as is
the case for several series in the {\em large $\b{0}$ limit}. This
makes it still possible to define a Borel sum, by choosing a
certain prescription for the Borel integral. At large energies,
the total result for a physical quantity is then supposed to
consist of a Borel sum of its perturbation series, combined with
some OPE power corrections, originating from certain values of the
non-perturbative condensates.

For a large class of observables, the dimension four gluon
condensate is the lowest dimension condensate in the OPE
\footnote{We consider massless QCD, so the quark condensate does
not contribute to the OPE.}. So the validity of the perturbative
expansion for a physical quantity, depending on one external
momentum $Q$, is intrinsically limited by a $Q^{-4}$ term.
However, the divergent perturbation series gives a stronger limit,
coming from the first UV renormalon, positioned at $-1/\b{0}$ in
the Borel plane. (Incidentally, for the definition of the Borel
tranform we use in section 2, the first UV renormalon actually
lies at -1.) Dealing with the series as an asymptotic series, one
concludes that, at the 'best' order of truncation, the
perturbative prediction can be trusted only up to a $Q^{-2}$ term.
In \cite{bz92}, the renormalization scale dependence of the
$Q^{-2}$ uncertainty was analyzed, in the large $\b{0}$ limit of
QCD. It was found that for large orders, the size of this term,
can be suppressed by increasing the scale, or equivalently,
decreasing the coupling constant in the perturbative expansion.

In this paper we will will extend this analysis, by allowing the
renormalization scale to vary with the order of truncation.  We
will show that if one lets $k$ $(\sim\ln\mu^2)$, grow linearly
with the order of truncation $N$, the {\em divergent} perturbation
series is turned into an expansion that {\em converges} to the
Borel integral, cut off at $N/k\equiv 1/\chi$, at least if
$\chi>1/2$. This allows us to formulate an expansion that
converges to the Borel sum up to a term $\propto Q^{-4}$, roughly
compatible with the dimension four gluoncondensate.

In section \ref{largek}, we examine the large order behavior for
the expansion of a generic QCD observable with the typical
singularity structure in the Borel plane. We arrive at an
asymptotic formula (\ref{master}), describing the large order
truncations for general values of $\chi$. This formula is then
applied to different toy expansions in section \ref{applications}.
First we consider the hypothetical situation, were one has only UV
renormalons, then we include the IR renormalons. Furthermore we
will use formula (\ref{master}) to establish formulations of the
{\em principle of minimal sensitivity} (PMS) and the {\em fastest
apparent convergence criterion} (FACC) that lead to a convergent
expansion.

\section{The 'scale' dependence at large orders}\label{largek}
Let us start by writing the general expression of the series
expansion for a physical quantity $\mathcal{R}$, depending on one
external scale $Q$, \be
\mathcal{R}(a_0)=a_0(r_0+r_1a_0+r_2a_0^2+\ldots)\,,\label{R}\ee
with $a_0\equiv \b{0}\a_s(Q)$ ($\approx 1/\ln(Q^2/\Lambda^2)$ for
large $Q$). We will examine how the expansion behaves, if we
reexpand in some new coupling $a(k)$: \be a(k)=\f{a_0}{1+k
a_0}\label{a(k)}\,.\ee In the large $\b{0}$ limit, this variation
corresponds to a change in the renormalization scale (RS)
$Q\rightarrow Qe^{k/2}$.  Of course nothing prevents us from
considering the same variation for real QCD expansions; but it
then no longer corresponds to a change of the RS.

The truncation $[\mathcal{R}(a_0)]_{N+1}\equiv
a_0(r_0+r_1a_0+\ldots+r_Na_0^N)$, is conveniently rewritten as a
Borel integral: \be
\int_{0}^{\infty}dt\,e^{-\f{t}{a_0}}[F(t)]_{N}\,, \label{Borel}\ee
with \be [F(t)]_{N}\equiv
r_0+r_1t+r_2\f{t^2}{2}+\ldots+r_N\f{t^N}{N!}\,.\label{F_N} \ee
Under the variation (\ref{a(k)}) the truncated expansion becomes
\be\label{R(ak)}[\mathcal{R}(a(k))]_{N+1}=a(k)[r_0(k)+r_1(k)a(k)+\ldots+r_N(k)a(k)^N]=
\int_{0}^{\infty}dt\,e^{-\f{t}{a(k)}}[e^{kt}F(t)]_{N}\,,\ee with
$[e^{kt}F(t)]_{N}$ the obvious generalization of (\ref{F_N}). One
now generally assumes that the series $r_0+r_1t+r_2t^2/2+\ldots$
has a finite (nonzero) radius of convergence, therefore it defines
a function $F(t)$, which is analytical around the origin. In fact,
it is established, through the use of several techniques, that
$F(t)$ has several types of singularities on the real axis
\cite{th77,p78,m85,b98} at a nonzero distance of the origin: on
the negative axis one finds the UV renormalons at the points
$-1,-2,\ldots$; on the positive axis there are IR renormalons at
the points $(1),2,\ldots$ and {\em instanton-anti instanton}
singularities at the points $4\b{0}\pi,8\b{0}\pi,\ldots$. With the
number of active flavors being \mbox{3-6
($\b{0}=(33-2N_f)/12\pi$)}, the singularities near $t=0$ are all
of the renormalon type. In the following we will consider the
situation described in the introduction, with no IR renormalon at
$t=1$.

One can now make use of the Cauchy integral theorem for analytical
functions \cite{aw95}, to write $[e^{kt}F(t)]_{N}$ as a
contourintegral along the contour $\mathcal{C}_0$ around the
origin, excluding the point $t$ and the singularities on the real
axis (see Figure 1): \bea [e^{kt}F(t)]_{N}&=&\sum_{n=0}^N
(e^{kt}F(t))^{(n)}|_{t=0}\f{t^n}{n!}\nonumber \\
&=&\sum_{n=0}^N \f{1}{2\pi i}\int_{\mathcal{C}_0}\!\!\!dz\,
e^{kz}F(z)
\f{t^n}{z^{n+1}}\nonumber\\
&=&\f{1}{2\pi
i}\int_{\mathcal{C}_0}\!\!\!dz\, e^{kz}F(z)\f{1-(\f{t}{z})^{N+1}}{z-t}\nonumber\\
&=&-\f{1}{2\pi
i}\int_{\mathcal{C}_0}\!\!\!dz\,\f{e^{kz}F(z)}{z-t}(\f{t}{z})^{N+1}\,.\label{C0}\eea
For the last equality, we used the analyticity of
$e^{kz}F(z)/(z-t)$ inside $\mathcal{C}_0$.
\begin{figure}
\begin{center}
\includegraphics[width=7cm]{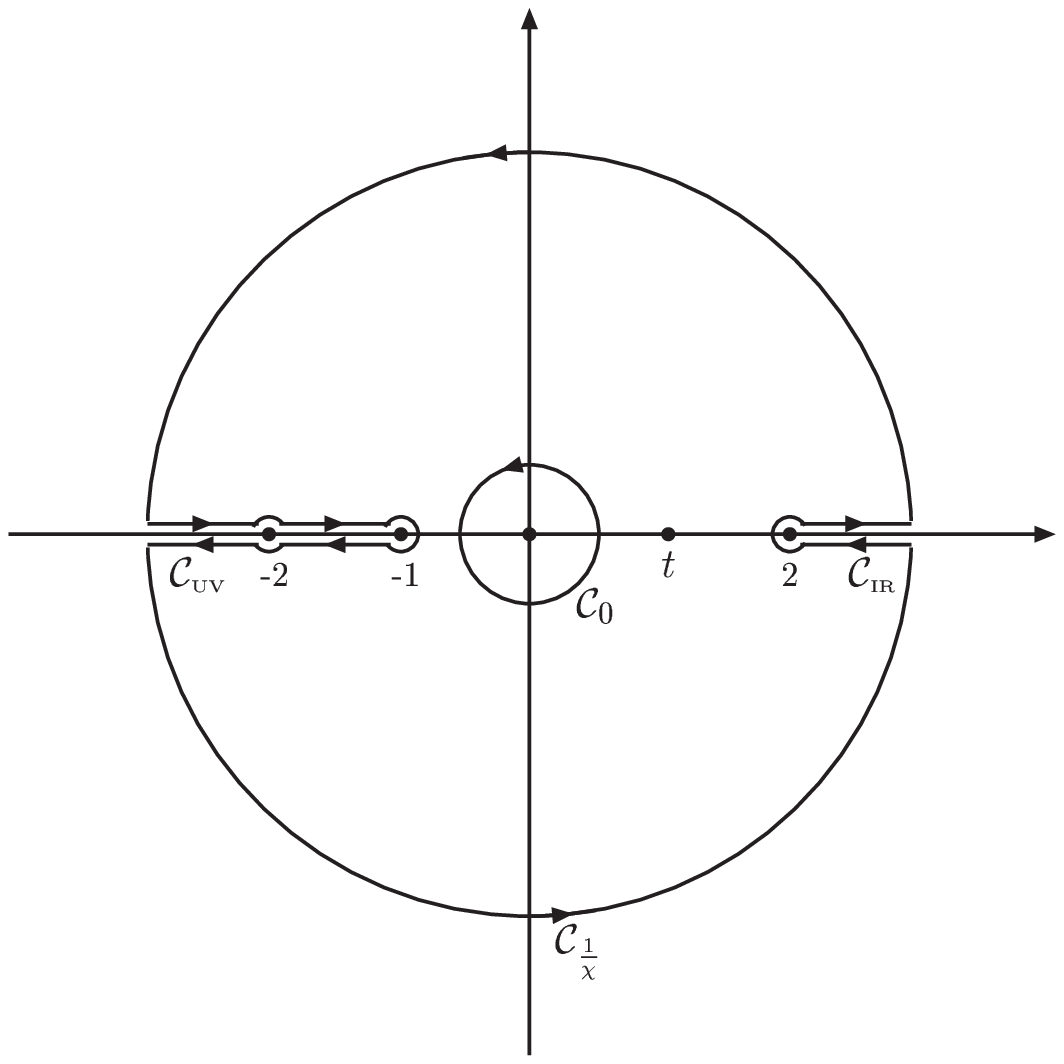}
\end{center}
\caption{The contour $\mathcal{C}_0$, equivalent to the contours
$\mathcal{C}_{1/\chi}$, $\mathcal{C}_{UV}$ and $\mathcal{C}_{IR}$,
if the residu at $t<1/\chi$ is picked up.}
\end{figure}

We are interested in the asymptotic behavior of (\ref{C0}); if we
put $k=N\chi$, the integrandum can be cast in the form
$(\ldots)\times e^{N(\chi z-\ln z)}$, which calls for the
saddlepoint technique \cite{aw95}. So we deform $\mathcal{C}_0$ to
a circle $\mathcal{C}_{1/\chi}$ with radius $1/\chi$, appropriate
for a saddlepoint evaluation.  If $t$ is lying inside the circle,
one has to pick up the residu $e^{kt}F(t)$, coming from the factor
$1/(z-t)$ in the integrandum. As shown in the figure, for $1/\chi$
larger than one (two), one also has to add the contour
$\mathcal{C}_{UV}$ ($\mathcal{C}_{IR}$), in order to exclude the
singularities and branch lines.  For simplicity, we will limit
ourselves in the following to pole singularities (without branch
lines) at the points $z_i=2,3\ldots$ on the positive axis and at
the points $z'_i=-1,-2\ldots$ on the negative axis, arriving at
\bea
\label{COCL}[e^{kt}F(t)]_{N}&=&\theta(\f{1}{\chi}-t)e^{kt}F(t)-\f{1}{2\pi
i}t^{N+1}\int_{\mathcal{C}_{1/\chi}}\!\!\!\f{dz}{z}\,\f{e^{N(\chi
z-\ln
z)}}{z-t}F(z)\nonumber\\&+&\sum_{z_i}\theta(\f{1}{\chi}-z_i)t^{N+1}\f{e^{N(z_i\chi-\ln
z_i)}}{z_i(z_i-t)}Res_{z=z_i}[F(z)]
\\
&+&\sum_{z'_i}\theta(\f{1}{\chi}-|z'_i|)(-1)^N
t^{N+1}\f{e^{-N(|z'_i|\chi+\ln|z'_i|)
}}{z'_i(z'_i-t)}Res_{z=z'_i}[F(z)] \,,\nonumber\eea with
$\theta(x)$ the unit stepfunction.

One can check that the singularities at $t=z_i$ of the term
\mbox{$\theta(1/\chi-t)e^{kt}F(t)$}, are exactly cancelled by the
corresponding terms in the summation over the singularities at the
positive axis. Therefore, when performing the $t$ integration of
the separate pieces in (\ref{COCL}), one can use any prescription,
as long as the same prescription is used for every piece. We will
use the Cauchy principal value prescription, defined as:\be
PV\!\!\!\int_0^\infty=\f{1}{2}(\int_0^{\infty+i\epsilon}+\int_0^{\infty-i\epsilon})\,.\ee

Before performing the $t$ integration, let us first work out the
$z$ integration in the second term of (\ref{COCL}). At the saddle
point, $z=1/\chi$, one has a a zero derivative for $N(\chi z-\ln
z)$. The second derivative is $+N\chi^{2}$, so the path of
steepest descent crosses the saddle point orthogonal to the real
axis. Furthermore for $z=e^{i\theta}/\chi$,\bea Re(N(\chi z-\ln
z))=N(\cos\theta+\ln\chi)\leq N(1+\ln \chi)\,,\eea thus
$\mathcal{C}_{1/\chi}$ is indeed an appropriate choice for the
saddle point contour. The integral is conveniently written as:
\bea-\f{1}{2\pi}t^{N+1}\chi
e^{N(1+\ln\chi)}\int^{+\pi}_{-\pi}\!\!\!d\theta\,\f{e^{N(e^{i\theta}-1-i\theta)}}{(1-\chi
t)+(e^{i\theta}-1)}\,F\Big(\f{1}{\chi}+\f{1}{\chi}(e^{i\theta}-1)\Big)
\,\label{CL}. \eea

With the reparametrization, $\theta=(2/N)^{1/2}x$, this expression
is readily expanded in powers of $1/N^{1/2}$. However, one should
be cautious for $t\approx 1/\chi$: in that case, the denominator
in (\ref{CL}) can not be expanded as \be \f{1}{1-\chi
t}(1-i\sqrt{\f{2}{N}}\f{x}{1-\chi t}+\ldots)\,. \ee  Later on we
will see that the dominant contribution of the $t$ integration of
(\ref{CL})  in (\ref{R(ak)}), comes exactly from this critical
region, therefore we have treated $\Delta\equiv (N/2)^{1/2}(\chi
t-1)$, as an order 1 term in the expansion of (\ref{CL}), arriving
at:\be \f{1}{2\pi}t^{N+1}e^{N(1+\ln \chi)}\chi
F(\f{1}{\chi})\Big(f_0(\Delta)+\sqrt{\f{2}{N}}f_1(\Delta)+\mathcal{O}(\f{1}{N})\Big)\,,\label{saddle}\ee
with \bea f_{0}(\Delta)&\equiv& i\int^{+\infty}_{-\infty}\!\!dx
\f{e^{-x^2}}{x+i\Delta}\,,\\
f_{1}(\Delta)&\equiv& \int^{+\infty}_{-\infty}\!\!dx\Big[
\f{e^{-x^2}}{x+i\Delta}\Big(\f{x^3}{3}-x\f{F'(\f{1}{\chi})}{\chi
F(\f{1}{\chi})}\Big)+\f{e^{-x^2}}{(x+i\Delta)^2}\f{x^2}{2}\Big]\,.\eea
One can verify that the odd function $f_0(\Delta)$ has a
discontinuity at $\Delta=0$ ($t=1/\chi$), $f_0(0\pm\epsilon)=\pm
\pi$, exactly compensating the discontinuity of the term
$\theta(\chi^{-1}-t)e^{kt}F(t)$ in (\ref{COCL}). Obviously the
discontinuities coming from the other step functions
$\theta(\chi^{-1}-z_i/|z'_i|)$ are also compensated by the
integration along $\mathcal{C}_{1/\chi}$. In principle one could
write \be
F(z)=\prod_{z_i}\f{1}{z-z_i}\prod_{z'_i}\f{1}{z-z'_i}\tilde{F}(z)\ee
and incorporate these singularities in the same way as we did for
the singularity at $z=t$. This complicates the calculations and
only modifies formula (\ref{saddle}) for $\chi^{-1}\approx
z_i/|z'_i| + \mathcal{O}(1/N^{1/2})$, effectively spreading the
discontinuous unit step of $\theta(\chi^{-1}-z_i/|z'_i|)$ in
(\ref{COCL}), over a region $\chi^{-1}=z_i/|z'_i| \pm
\mathcal{O}(1/N^{1/2})$.

We are now ready to perform the $t$ integration of (\ref{COCL}) in
(\ref{R(ak)}). First, we will consider the integration over the
saddle point contribution (\ref{saddle}). Notice that the
integrandum can be cast in the form $(\ldots)\times e^{N(\ln
t-\chi t)}$, with the opposite power of the exponent as for the
$z$-integration, so again we have a saddle point at $1/\chi$, but
now with the positive axis as the path of steepest descent. With
the reparametrization \be t=\f{1}{\chi} +
\sqrt{\f{2}{N}}\f{x}{\chi}\,,\label{t(x)}\ee the integral reduces
to: \bea && \f{e^{-\f{1}{\chi a_{0}}}F(\f{1}{\chi})}{2\pi
\chi}\sqrt{\f{2}{N}}\int^{+\infty}_{-(N/2)^{1/2}}\hspace{-1.1cm}dx\hspace{0.1cm}
e^{-\f{1}{\chi a_{0}}\sqrt{\f{2}{N}}x-\sqrt{2N}x
+N\ln(1+\sqrt{\f{2}{N}}x)}(1+\sqrt{\f{2}{N}}x)\Big[f_{0}(x)+\sqrt{\f{2}{N}}f_1(x)+\mathcal{O}(\f{1}{N})\Big]\nonumber\\
&\approx& \f{e^{-\f{1}{\chi a_{0}}}F(\f{1}{\chi})}{\pi
\chi}\f{2}{N}\,\,\int^{+\infty}_{0}\hspace{-0.5cm}dx
e^{-x^2}\Big(f_0(x)\big[x(1-\f{1}{\chi
a_0})+\f{2}{3}x^3\big]+f_1(x)\Big)\,\,\Big(1+\mathcal{O}(\f{1}{N^{1/2}})\Big)\,.\label{CLT}
\eea In the last step we used $f_0(-x)=-f_0(x)$ and
$f_1(-x)=f_1(x)$. With the introduction of some auxiliary
integrations, the integral is calculated exactly, arriving at \bea
\f{1}{N}\f{e^{-\f{1}{\chi a_{0}}}}{\chi}\Big((1-\f{1}{2\chi
a_0})F(\f{1}{\chi})-\f{F'(\f{1}{\chi})}{2\chi}\Big)\Big(1+\mathcal{O}(\f{1}{N^{1/2}})\Big)\,\label{approxt1}.
\eea

Each term in (\ref{COCL}), coming from a renormalon or
instanton-anti-instanton at $z_i$ ($z'_i$), contributes a term \be
(\ldots)\times\int^{+\infty}_{0}\hspace{-0.5cm}dt\,\,e^{-t(\f{1}{a_0}+N\chi)}\f{t^{N+1}}{z_i-t}\,
\ee to the $t$ integration (\ref{R(ak)}). With the same
reparametrization (\ref{t(x)}), this is easily expanded (for
$1/\chi\approx\hspace{-0.3cm}/\,\, z_i$) as: \be (\ldots)\times
e^{-\f{1}{\chi a_0}-N(1+\ln\chi)}\f{1}{\chi(\chi
z_i-1)}\sqrt{\f{2\pi}{N}}\Big(1+\mathcal{O}(\f{1}{N})\Big)\,.\label{approxt2}\ee

After collection of all the terms, we finally arrive at the
asymptotic formula for the truncated expansion in $a(k)$
($k=N\chi$), of an observable $\mathcal{R}(a_0)$ with Borel
transform $F(t)$:

\bea
[\mathcal{R}(a(k))]_{N+1}&=&PV\!\!\int^{\f{1}{\chi}}_{0}\hspace{-0.2cm}dt\,\,e^{-\f{t}{a_0}}F(t)\,+\,\f{1}{N}\f{e^{-\f{1}{\chi
a_{0}}}}{\chi}\Big((1-\f{1}{2\chi
a_0})F(\f{1}{\chi})-\f{F'(\f{1}{\chi})}{2\chi}\Big)\Big(1+\mathcal{O}(\f{1}{N^{1/2}})\Big)\nonumber\\
&&+\theta(\f{1}{\chi}-2)\,e^{-\f{1}{\chi
a_0}}Res_{z=2}[F(z)]\,\,\f{e^{N(2\chi-1-\ln(2\chi))}}{2\chi(2\chi-1)}\sqrt{\f{2\pi}{N}}\Big(1+\mathcal{O}(\f{1}{N})\Big)\label{master}\\
&&+\theta(\f{1}{\chi}-1)\,(-1)^Ne^{-\f{1}{\chi
a_0}}Res_{z=-1}[F(z)]\,\,\f{e^{-N(\chi+1+\ln\chi)}}{\chi(\chi+1)}\sqrt{\f{2\pi}{N}}\Big(1+\mathcal{O}(\f{1}{N})\Big)\,.\nonumber\eea
We have only kept the contribution of the singularity closest to
the origin on the positive/negative axis at the point $2/-1$; for
all values of $\chi$, this singularity dominates the contributions
of the other singularities on the positive/negative axis by a
factor $c^N$, with $c>1$.

From the discussion on the validity of (\ref{saddle}), we know
that the asymptotic formula is invalid at the 'steps' of the
$\theta$ functions, $\chi^{-1}=2/-1\pm \mathcal{O}(1/N^{1/2})$.

Also note that the approximations (\ref{approxt1}) and
(\ref{approxt2}), for the $t$ integration are only valid for
\mbox{$N\gg 1/(\chi a_0)$}. For the relevant values of $\chi$ and
$a_0$ that we will use later on, this means $N\gg 10$. Therefore,
the formula (\ref{master}) can certainly not be used to examine
for instance the NNLO ($N=2$) result. It should be treated as an
{\em asymptotic} formula, valid for (very) large $N$, assessing
the convergence/divergence of the truncations in the limit
$N\rightarrow \infty$ for a certain value of $\chi$.

We end this section by quoting the result for the asymptotic
formula, in the case of a Borel transform with general
singularities. One can verify that the renormalon contributions
coming from the $z$ integrations along $\mathcal{C}_{IR}$ and
$\mathcal{C}_{UV}$ change to:\bea
&&\theta(\f{1}{\chi}-2)\,e^{-\f{1}{\chi
a_0}}K\f{\sin(\pi b)}{\pi}\Gamma(-b)\,N^b(\f{1}{2}-\chi)^b\f{e^{N(2\chi-1-\ln(2\chi))}}{2\chi(2\chi-1)}\sqrt{\f{2\pi}{N}}\Big(1+\mathcal{O}(\f{1}{N})\Big)\label{masterII}\\
&&+\theta(\f{1}{\chi}-1)\,(-1)^{N+1}e^{-\f{1}{\chi
a_0}}K'\f{\sin(\pi
b')}{\pi}\Gamma(-b')\,N^{b'}(1+\chi)^{b'}\,\,\f{e^{-N(\chi+1+\ln\chi)}}{\chi(\chi+1)}\sqrt{\f{2\pi}{N}}\Big(1+\mathcal{O}(\f{1}{N})\Big)\,,\nonumber\eea
for the singular behavior,

\bea F(z)\stackrel{z \rightarrow
2}{=}\f{K}{(2-z)^{1+b}}\,\,,\hspace{1cm}F(z)&\stackrel{z
\rightarrow -1}{=}&\f{K'}{(1+z)^{1+b'}}\,.\eea Notice that the
exponentials, responsible for the large order behavior, have the
same arguments as in (\ref{master}). For $\chi>1/2$, the first two
terms in (\ref{master}) stay the same, and we have essentially the
same features for the $\chi$ dependent large order behavior as in
the case of pole singularities.
\section{Applications}\label{applications}
\subsection{The UV renormalons}

It is instructive to consider first the fictitious case of a
series that has no IR renormalons at all. In that case the IR
renormalon term in the asymptotic formula (\ref{master}) should be
left out, leaving us with the UV renormalon term as the only
possible source of divergence. For $\chi<\chi_\star$, with
$\chi_\star\approx 0.28$ the solution of \be \chi+1+\ln\chi=0\,,
\ee this UV term diverges exponentially at large orders while it
it disappears exponentially for $\chi>\chi_\star$. So, only in the
latter case the expansion is convergent: \be
[\mathcal{R}(a(N\chi))]_{N+1}\stackrel{N\rightarrow
\infty}{=}\!\!\int^{\f{1}{\chi}}_{0}\hspace{-0.2cm}dt\,\,e^{-\f{t}{a_0}}F(t)\,.\ee
\begin{figure}
\begin{tabular}{cc}

\includegraphics[width=7.5cm]{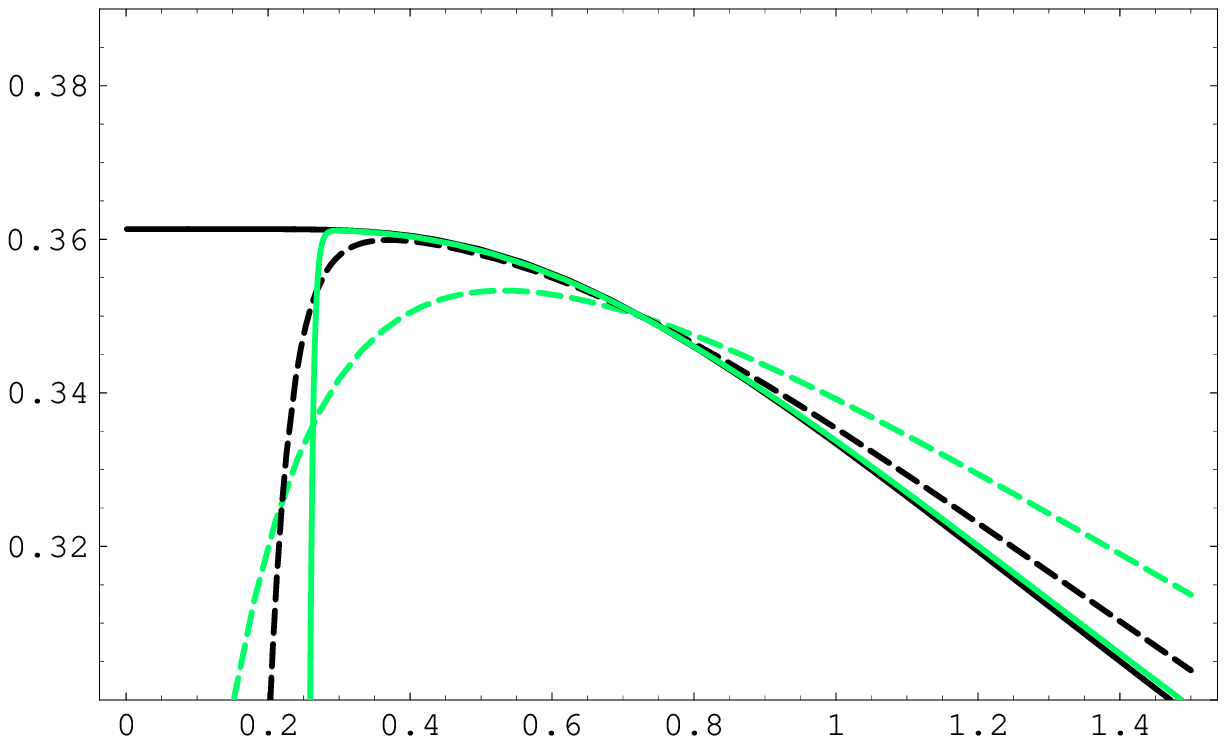}&\includegraphics[width=7.5cm]{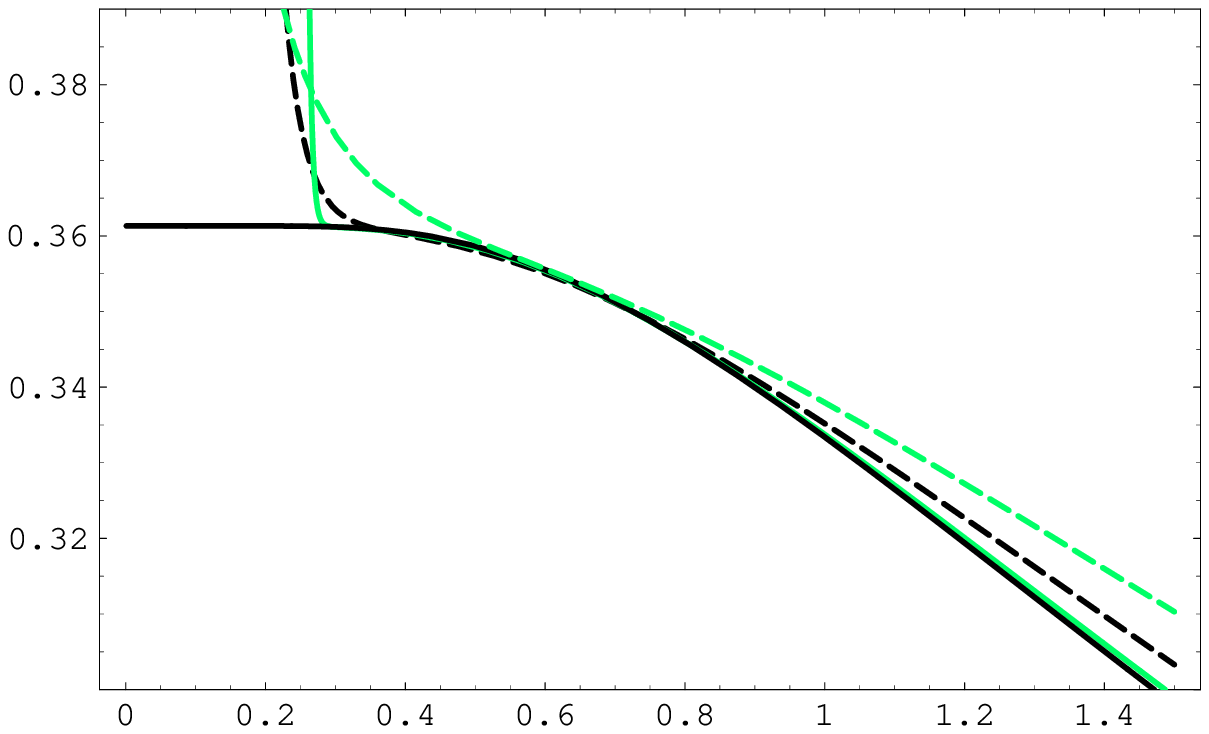}

\end{tabular}
\caption{The truncations ${[\mathcal{R}(a(N\chi))]_{N+1}}$, for
$N=$3[grey dashed], 9 [dashed], 49[grey] (left) and $N=$4, 10, 50
(right) of the series with Borel transform $F(t)=1/(1+t)$, at
$a_0=1/2$, as a function of $\chi$. Also plotted: the Borel
integral cut off at $1/\chi$ [full line]. }
\end{figure}

This is illustrated in Figure 2, where we plot some approximants
$[\mathcal{R}(a(N\chi))]_{N+1}$ as a function of $\chi$, at
$a_0=1/2$ for the Borel transform $F(t)=1/(1+t)$. This specific
model was examined some time ago by Stevenson \cite{s84} in the
context of the PMS. One clearly observes the behavior predicted by
(\ref{master}): for $\chi>\chi_\star$ the expansion converges
towards the Borel integral cut off at $1/\chi$, for
$\chi<\chi_\star$ it diverges. The alternation of the divergence
originates from the factor $(-1)^N$ in the UV term of the
asymptotic formula (\ref{master}). At the transition region
$\chi\approx \chi_\star$, one has an extremum for odd orders $N$
and an inflection point for even orders.

From (\ref{master}) one easily shows for a general $F(t)$ (with no
IR renormalons), that for large odd orders an extremum occurs at:
\be
\tilde{k}(N)=N\tilde{\chi}\stackrel{N\rightarrow\infty}{=}N\chi_{\star}+\f{\f{1}{2}\ln
N}{1+1/\chi_{\star}} \,,\label{kms} \ee that is if
$Res_{z=-1}[F(z)]/F(1/\chi_{\star})>0$; in the opposite case the
extremum occurs for even orders. Furthermore, the substitution of
(\ref{kms}) in (\ref{master}) shows that, although
$\tilde{\chi}\rightarrow \chi_ \star$, the expansion with $k$ at
the extremum $\tilde{k}$ still converges:\bea
[\mathcal{R}(a(\tilde{k}))]_{N+1}&\stackrel{N\rightarrow\infty}{=}&
\!\!\int^{\f{1}{\chi_{\star}}}_{0}\hspace{-0.2cm}dt\,\,e^{-\f{t}{a_0}}F(t)\,,
\eea in agreement with \cite{s84} for $F(t)=1/(1+t)$, also
confirmed in figure 2.


With $Res_{z=-1}[F(z)]/F(1/\chi_{\star})>0$ ($<0$), there is no
extremum at $k\approx N\chi_\star$, for large even (odd) orders.
If

\be
\f{1}{a_0}>\f{F'(\f{1}{\chi_\star})}{F(\f{1}{\chi_\star})}+2\chi_{\star}\,,\label{conda0}\ee
formula (\ref{master}) instead predicts a zero of the second
derivative at: \be
\bar{k}(N)=N\bar{\chi}\stackrel{N\rightarrow\infty}{=}N\chi_{\star}+\f{\f{3}{2}\ln
N}{1+1/\chi_{\star}}\,. \ee Generally, the condition
(\ref{conda0}) is fulfilled for $a_0$ small enough; for
$F(t)=1/(1+t)$, it is met for $a_0<2.9$. Again, we have
 \bea
[\mathcal{R}(a(\bar{k}))]_{N+1}&\stackrel{N\rightarrow\infty}{=}&
\!\!\int^{\f{1}{\chi_{\star}}}_{0}\hspace{-0.2cm}dt\,\,e^{-\f{t}{a_0}}F(t)\,,\eea
so the PMS converges to the same value both for odd and even
orders, at least if we take $k_{PMS}$ at the extremum/inflection
point in the neighborhood of $N\chi_{\star}$. This is in
disagreement with the analysis of Stevenson \cite{s84} on
$F(t)=1/(1+t)$; he predicted the PMS to be biconvergent, with an
extra term $\propto e^{-1/(\chi_\star a_0)}$, for even orders.

In the QCD phenomenology one is generally interested in the Borel
sum, i.e. the Borel integral cut off at infinity (with a certain
prescription in the case of singularities at the positive $t$
axis). Ordinary perturbation theory, with a fixed value of $k$,
gives a divergent expansion with an error $\propto
e^{-1/a_0}\approx Q^{-2}$, at the 'best' order. For $a_0$ not too
large, we can estimate \be
\int^{\f{1}{\chi_{\star}}}_{0}\hspace{-0.2cm}dt\,\,e^{-\f{t}{a_0}}F(t)\approx
\int^{\infty}_{0}\hspace{-0.2cm}dt\,\,e^{-\f{t}{a_0}}F(t) +
\mathcal{O}(e^{-\f{1}{\chi_\star a_0}})\,.\ee Thus our results
imply that, with an appropriate choice of $k$, in the absence of
IR renormalons, one can approach the Borel sum much better, up to
a correction $\propto Q^{-2/\chi_\star}\approx Q^{-7}$. One way to
achieve this is to take, for each order, $k$ equal to
$N\chi_\star$. (It can be checked from (\ref{master}) that this
choice for $k$, still gives a convergent expansion.) Another way
is to apply the PMS, as formulated in this subsection.
\subsection{Including the IR renormalons} Things change when the
IR renormalons come into play. Now, the IR renormalon term in
(\ref{master}) diverges exponentially for $\chi<1/2$; so the
expansion only converges for $\chi>1/2$.
\begin{figure}
\begin{tabular}{cc}

\includegraphics[width=7.5cm]{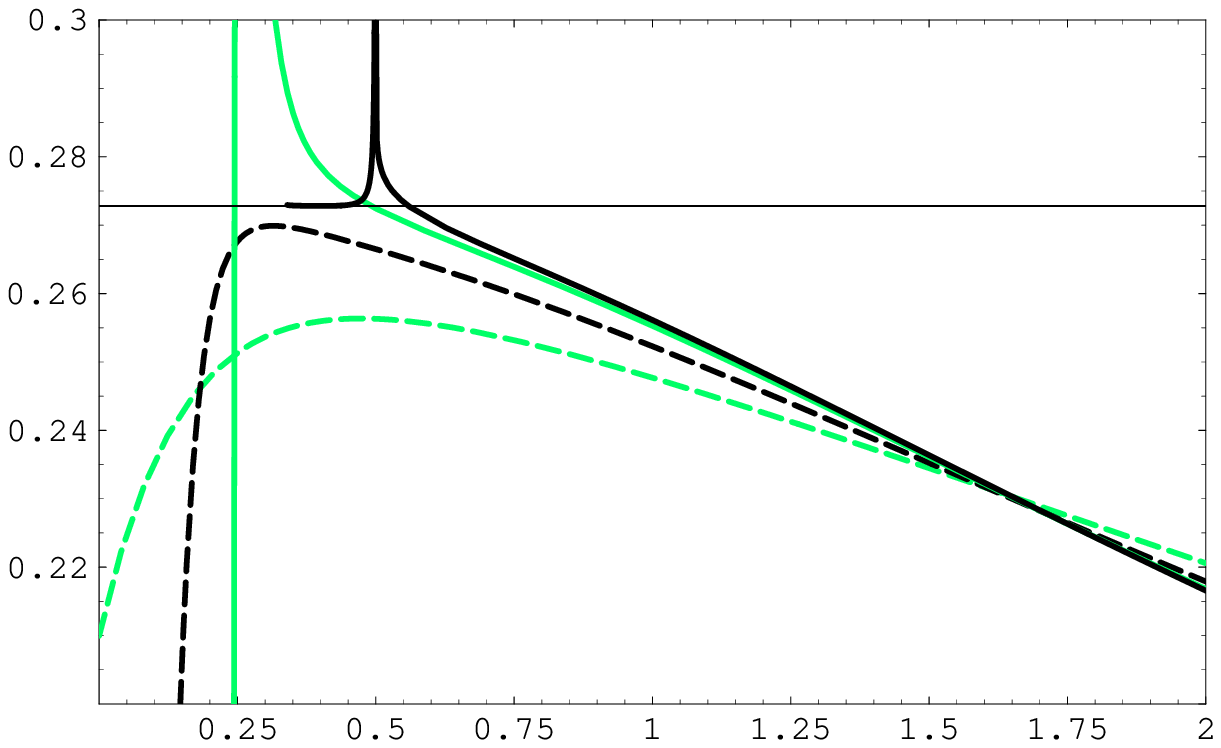}&\includegraphics[width=7.5cm]{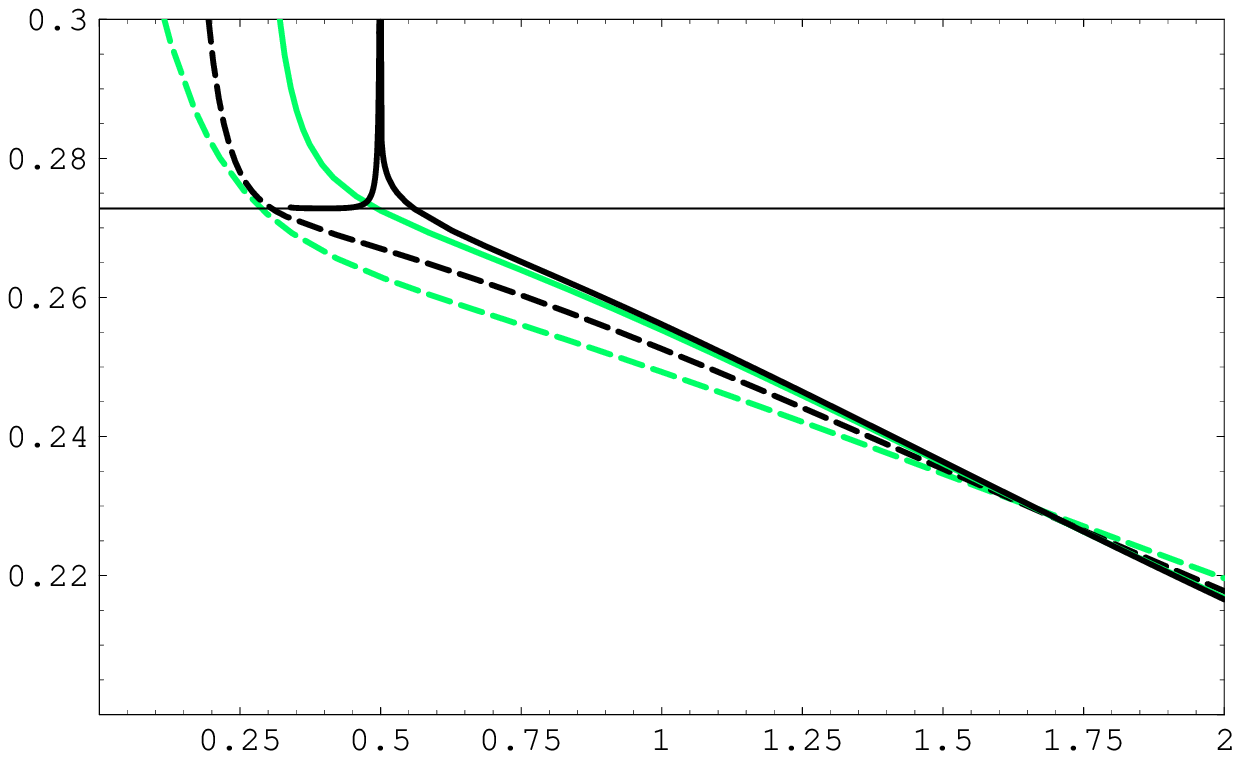}

\end{tabular}
\caption{The approximants ${[D(a(N\chi))]_{N+1}}$ of the large
$\b{0}$ Adler D-function, for \mbox{$N=$3[grey dashed]}, 9
[dashed], 49[grey] (left) and $N=$4, 10, 50 (right), at $a_0=0.3$,
as a function of $\chi$. Also plotted: the Borel integral cut off
at $1/\chi$ [full line] and the Borel sum [small full line].}
\end{figure}
We have performed explicit calculations on the large $\b{0}$ limit
of the Adler D-function. In this limit the Borel transform is
calculated exactly\cite{b93}:\be
F_D(t)=\f{32}{3}\f{1}{2-t}\sum^\infty_{p=2}
\f{(-1)^pp}{[p^2-(1-t)^2]^2}\,,\ee with, as expected,
singularities for $t=-1,-2,\ldots$ and $t=2,3,\ldots$. Also note
that the Borel integral converges at infinity, for $a_0>0$. In
order to keep the calculation time under control, we have
truncated the $p$-summation at $p=100$. In Figure 3, we plot
several approximants $[D(a(N\chi))]_{N+1}$, as a function of
$\chi$, at $a_0=0.3$, together with the Borel integral cut off at
$1/\chi$ and the Borel sum. We clearly observe the convergence to
the cut Borel integral for $\chi>1/2$ and the divergence for
$\chi<1/2$. At the latter region, there is a turnover for odd
orders.  This occurs when the UV renormalon term starts dominating
the IR renormalon term in (\ref{master}) at $\chi=\ln2/3\approx
0.23$ for large orders.

Due to the singularity of $F_{D}(t)$ at $t=2$, the cut Borel
integral itself also diverges at $\chi\approx 1/2$. (For the sake
of clarity, we have only plotted the Borel integral for
$\chi>1/3$, thereby omitting all the other divergencies at
$\chi=1/3,1/4,\ldots$.) Therefore, the best value for $\chi$, in
the sense that the corresponding expansion approaches the Borel
sum as close as possible, no longer occurs exactly at the
transition between convergence and divergence of the expansion.
Instead, it is recommended to take $\chi$ slightly larger than
1/2. If $a_0$ is not too large, a good choice will be \be
\tilde{\chi}=1/(2-ca_0+\mathcal{O}(a_0^2))\,,\label{chitilde }\ee
where $c$ is some positive constant. Indeed, one easily finds that
for a general singularity (with $b>-1$) \be F(t)\stackrel{t
\rightarrow 2}{=}\f{K}{(2-t)^{1+b}}\,,\label{sing}\ee the Borel
integral cut at $1/\tilde{\chi}$ is approximated as: \be
\int^{1/\tilde{\chi}}_0\!\!\!dt\, e^{-\f{t}{\a_0}}F(t)\approx
PV\!\!\int^{\infty}_0\!\!\!dt\,
e^{-\f{t}{\a_0}}F(t)+e^{-\f{2}{a_0}}\Big[a_0^{-b}K[
C+\mathcal{O}(a_0)]+\mathcal{O}(a_0)\Big]
[1+\mathcal{O}(e^{-\f{1}{a_0}})]\,,\label{approx}\ee with $C$
depending on $c$: \be C=
PV\!\!\!\int^{\infty}_{-c}\hspace{-0.2cm}dt\hspace{0.2cm}\f{e^{-t}}{(-t)^{1+b}}\,.\label{C}\ee
So we find that, with an appropriate choice for $k$, one can
approach the Borel sum up to an error $\propto e^{-2/a_0}\propto
Q^{-4}$, whereas the ordinary perturbation series, with fixed $k$,
gives a minimal error $Q^{-2}$. In addition, if we restrict
ourselves in the OPE to the lowest dimension condensate, and to
the lowest order term of the Wilsoncoefficient, the $Q^{-4}$ error
can be compensated by a redefinition of the condensate. Indeed,
with the singular behavior (\ref{sing}), the OPE power correction
(proportional to the ambiguity of the Borel integral) to the Borel
sum, is written as: \be G
e^{-\f{2}{a_0}}a_0^{-b}K[1+\mathcal{O}(a_0)]\,,\ee where $G$
parameterizes the experimentally fitted, non-perturbative
condensate. Thus, at leading order in $a_0$, the $Q^{-4}$ error
(\ref{approx}) is absorbed in the condensate by the replacement
$G\rightarrow G'=G-C$.

From the second column of Tables 1 and 2, one clearly observes
that the approximants with \mbox{$k=N/(2-2a_0)$} indeed converge
to the Borel integral cut off at $(2-2a_0)$, for the Adler
$D$-function at $a_0=0.2$ and $a_0=0.4$. This gives an error of
$-0.2\%$ and $-8\%$ with respect to the Borel sum for
$N\rightarrow \infty$, whereas the divergent series with $k$ fixed
at zero, has a minimal error of $5\%$ and $29\%$.  Furthermore,
there is also a considerable improvement at finite orders.

\begin{table}
\begin{center}
\begin{tabular}{|l|c|c|c|c|} \hline
$N\backslash k$&$0$&$N/(2-2a_0)$&$k_{PMS}$&$k_{FACC}$\\
\hline

1& -10.77& -7.62& -7.26$^{*}$& -7.26$^{*}$\\
2& 5.06& -1.31& -3.43&-2.34\\
3& -5.76& -1.95& -1.80$^{*}$& -1.80$^{*}$\\
4& 5.39& -1.26& -1.10& -0.93\\
5& -7.38& -1.12&  -0.75$^{*}$&  -0.75$^{*}$\\
7& -17.6& -0.83& -0.83$^{\diamond}$& -0.41$^{*}$\\
10& 150& -0.63&-0.24&-0.22\\
25&-6$\times10^{10}$&-0.35&-0.048&-0.0070\\
50& 8$\times10^{32}$&-0.26&-0.068&-0.034\\
100&5$\times10^{91}$&-0.21&-0.088&-0.055\\
$\infty$&$\pm\infty$&-0.16&-0.12&-0.095\\
\hline
\end{tabular}
\caption{Deviations in terms of percentage of the approximants
$[D(a(k))]_{N+1}$ for the Adler $D$-function at $a_0=0.2$, with
respect to the Borel sum, for different values of $k$. For the
$N\rightarrow \infty$ limits, we took the cut Borel integral, with
the cuts determined according to the text.}
\end{center}
\end{table}
\begin{table}
\begin{center}
\begin{tabular}{|l|c|c|c|c|}
\hline
$N\backslash k$&$0$&$N/(2-2a_0)$&$k_{PMS}$&$k_{FACC}$\\
\hline

1& -33.5& -21.7& -21.6$^{*}$& -21.6$^{*}$\\
2& 28.8& -12.1& /& -15.2\\
3& -56.3& -12.9& -17.3& -9.83$^{*}$\\
4& 119& -11.8& -16.1$^{\diamond}$& -8.73\\
5& -283& -11.4& -10.8& -4.96$^{*}$\\
7& -2685& -10.7& -11.5$^{\diamond}$& -1.94$^{*}$\\
10& 1.8$\times10^5$& -10.0& -14.1& -2.30\\
25&-2$\times10^{18}$&-8.89&/&-3.18\\
50& 9$\times10^{47}$&-8.40&/& -4.14\\
100&7$\times10^{121}$&-8.14&/&-4.71\\
$\infty$&$\pm \infty$&-7.85&/&-5.37\\ \hline
\end{tabular}
\caption{The same as in Table 2, now for $a_0=0.4$. }
\end{center}
\end{table}
As in the case without the IR renormalons, one can also formulate
some condition on the approximants $[\mathcal{R}(a(k))]_{N+1}$,
which selects for each order a certain value of $k$. This results
in a convergent behavior, if for large orders $k\rightarrow
N\chi$, with $\chi>1/2$. In this region, a condition on the
approximants, reduces for large $N$, to a condition on the Borel
integral cut off at $1/\chi$. One can for instance define a PMS
criterion as: \be \f{\delta^2[\mathcal{R}(a(k))]_{N+1}}{\delta
k^2}|_{k=k_{PMS}}=0\,,\label{PMS}\ee which results in the large
order behavior \be
[\mathcal{R}(a(k_{PMS}))]_{N+1}\stackrel{N\rightarrow\infty}{=}\int^{1/\chi_{PMS}}_0\!\!\!dt\,
e^{-\f{t}{\a_0}}F(t)\,,\label{limkPMS}\ee where \be
\f{\delta^2\Big(\int^{1/\chi}_0\!dt\,
e^{-\f{t}{\a_0}}F(t)\Big)}{\delta^2\chi}|_{\chi=\chi_{PMS}}\,=0
\Leftrightarrow\f{1}{a_0}=2\chi_{PMS}+\f{F'(1/\chi_{PMS})}{F(1/\chi_{PMS})}.\label{inflection}\ee
For small values of $a_0$, with the singular behavior
(\ref{sing}), one has
$\chi_{PMS}=1/(2-(1+b)a_0+\mathcal{O}(a_0^2))$, leading to a
$Q^{-4}$ error with respect to the Borel sum. However, Eq.
(\ref{inflection}) does not necessarily have a solution for all
values of $a_0$. This is the case for the Borel transform $F_D(t)$
of the Adler $D$-function when $a_0>0.34$. Evidently, one will
then no longer find an inflection point for large orders,
rendering the PMS criterion (\ref{PMS}) useless. Note that the
more common PMS criterion, with $k_{PMS}$ defined at an extremum,
will generally have no solutions at large even/odd
\footnote{depending on the relative sign of the IR and UV terms in
(\ref{master})} orders. Furthermore, taking $k$ at the extremum
for large odd/even orders will not result in a convergent
expansion, since this extremum occurs at $\chi\approx\ln2/3<1/2$,
as mentioned before.

We stress that our analysis is only valid for large $N$. It may
well happen that (\ref{PMS}) has no solution at certain finite
orders, although $\chi_{PMS}$ (\ref{inflection}) exists, implying
a solution for (\ref{PMS}) at large orders. Therefore, at finite
orders, the PMS should be applied with a certain amount of
flexibility. For instance, in the case of the Adler $D$-function
at $a_0=0.2$, we find that (\ref{PMS}), has no solution at the odd
orders $N\leq21$.  At the orders $N=1,3,5$, we have taken
$k_{PMS}$ at the extremum of $[D(a(k))]_{N+1}$, indicated by a $*$
in the third column of Table 1. At the orders $N=7,\ldots,21$,
$k_{PMS}$ was taken at the (non-zero) minimum of \be
|\f{\delta^2[D(a(k))]_{N+1}}{\delta k^2}|\,,\ee indicated by a
$\diamond$. (This minimum does not exist for $N<7$.) For large
orders, one can observe the convergence to the Borel integral cut
off at $1/\chi_{PMS}$. The improvement at finite orders is at
least equally large (except for $N=2$) as for the expansion with
$k=N/(2-2a_0)$.

For $a_0=0.4$,  we find that no value of $k$ can be identified as
$k_{PMS}$, when $N>16$, not even if we relax the condition, as we
did for the low odd orders at $a_0=0.2$. This is in agreement with
our analysis, since $\chi_{PMS}$ does not exist for $a_0=0.4$.

We have also considered a  FACC, defining $k_{FACC}$ at a minimum
of

\be
|\f{[\mathcal{R}(a(k))]_{N+1}-[\mathcal{R}(a(k))]_{N}}{[\mathcal{R}(a(k))]_{N+1}}|
=|\f{a(k)^{N+1}r_N(k)}{[\mathcal{R}(a(k))]_{N+1}}|\,.\label{FACC}\ee
From (\ref{master}) one easily shows that the FACC expansion
converges
 \be
[\mathcal{R}(a(k_{FACC}))]_{N+1}\stackrel{N\rightarrow\infty}{=}\int^{1/\chi_{FACC}}_0\!\!\!dt\,
e^{-\f{t}{\a_0}}F(t)\,,\label{limkFACC}\ee with $\chi_{FACC}$ at
the minimum of \be |\f{1}{\chi}\f{e^{-1/(\chi
a_0)}F(1/\chi)}{\int^{1/\chi}_0\!dt\, e^{-\f{t}{\a_0}}F(t)}|\,.\ee
For small values one finds again
$\chi_{FACC}=1/(2-(1+b)a_0+\mathcal{O}(a_0)^2)$, thus at leading
order in $a_0$, the $Q^{-4}$ correction equals the one for the PMS
expansion. One can show that (if $b>-1$ in (\ref{sing})) a
sufficient condition for $\chi_{FACC}(>1/2)$ to exist is:
$F'(0)<1/a_0$. For the Adler $D$-function, this condition is
fulfilled for any positive value of $a_0$, so in contrast with the
PMS, the FACC will now have a large order solution, both for
$a_0=0.2$ and $a_0=0.4$, as can be seen from the last column of
Tables 1 and 2. At large orders, we see the convergence of the
FACC expansion to the Borel integral cut at $1/\chi_{FACC}$. At
low orders, the improvement is generally larger than for the PMS
or for $k=N/(2-2a_0)$. At low odd orders ($N\leq23,\,11$ for
$a_0=0.2,\,0.4$) $k_{FACC}$ was taken at a zero of (\ref{FACC}).
This is indicated again by a $*$, because it is also an extremum
of $[D(a(k))]_{N+1}$, which follows from \be \f{\partial
[\mathcal{R}(a(k))]_{N+1}}{\partial k}=-(N+1)a(k)^{N+2}r_N(k)\,.
\ee

\section{Conclusions}

The main result of this paper is formula (\ref{master}),
describing the asymptotic behavior of truncations of the expansion
for a generic QCD observable with the typical singularity
structure in the Borel plane. We have successfully tested the
formula on several toy expansions. It implies that, with a  'good'
order dependent choice for the 'large $\b{0}$ renormalization
scale' $k$, one can recover a convergent expansion from the
divergent perturbation series. In the absence of IR renormalons,
one can make this expansion converge to a limit, differing from
the Borel sum by a term $\propto Q^{-7}$ at large $Q$. For the
realistic case, with IR renormalons starting at $t=2$ in the Borel
plane, one can obtain a difference $\propto Q^{-4}$, contrasting
the minimal $Q^{-2}$ error of the divergent ordinary (fixed $k$)
perturbation series. To achieve this one can simply take
$k=N/(2-ca_0)$, or, under certain conditions, one can apply some
self-consistent criterion like the PMS (\ref{PMS}) or the FACC
(\ref{FACC}). The latter criteria will generally converge faster.

Note that in the case of a first IR renormalon at $t=1$, a
straightforward generalization of (\ref{master}), will only show
convergence for $\chi>1$, implying a $Q^{-2}$ error, again in
accordance with the lowest dimension OPE condensate.

For our results to be relevant on a practical level, the
convergence has to set in fast enough. Although we generally find,
that $k=N/(2-ca_0),k=k_{PMS}$ or $k=k_{FACC}$ are also good
choices at $N=1,2$, we stress that one should not expect miracles
at these low orders. In our opinion, no procedure to determine the
{\em optimal} renormalization scale and scheme, will ever bypass
the simple fact that one is using only two or three of an infinite
set of coefficients. The main thing one should ask for is that the
procedure at least results in a convergent expansion, with a
finite $N\rightarrow \infty$ limit, that is (by definition)
recovered to arbitrary precision, by calculating enough orders.

Throughout the paper we have worked with perturbation series for
which it was possible to define a Borel sum, or equivalently, for
which the Borel integral converged at infinity. The success of the
expansions was evaluated by looking how close they approached the
Borel sum (+ some OPE power corrections). As mentioned in the
introduction, for real QCD, it is argued from general principles
\cite{th77}, that the Borel integral in fact diverges at infinity.
This obscures the picture of the non-perturbative power
corrections, naturally arising from the ambiguities of the Borel
sum, since the Borel sum itself is divergent for any prescription.
However, we can still use our results, to motivate heuristically
the need for a non-perturbative correction $\propto Q^{-4}$ rather
than $\propto Q^{-2}$ (for the first IR renormalon at 2). Indeed,
the $N\rightarrow \infty$ limit of the expansion can be
interpreted as an approximant $\mathcal{R}(\chi)$ for a physical
quantity $\mathcal{R}_0$, dependent on one unphysical parameter
$\chi$. It is then practice to estimate the error $\Delta$ as
proportional to small variations of the unphysical parameter. In
our case, this leads to the estimation \be\Delta(\chi)\propto
-\f{1}{\chi^2}F(\f{1}{\chi})e^{-1/(a_0\chi)}\propto Q^{-2/\chi}
\label{error}\,, \ee for large $Q$ and $\chi>1/2$. So the physical
quantity $\mathcal{R}_0$, seems to be approached as close as
possible, by taking $\chi\rightarrow 1/2$, leaving an intrinsic
'error', or non-perturbative correction, $\propto Q^{-4}$. To be
precise, one should take $\chi=1/(2-ca_0)$, otherwise the $Q$
independent term $F(1/\chi)/\chi^2$ in (\ref{error}) blows up. For
the generic singular behavior (\ref{sing}), we then have (for
$b>-1$): \be \mathcal{R}(c')=\mathcal{R}(c)+
e^{-\f{2}{a_0}}a_0^{-b}K\int^{c}_{c'}\!\!\!\!dt
\f{e^t}{t^{1+b}}\,\,\,\,+\,\,\ldots\,.\label{ambig}\ee With the
above-mentioned argument in mind, this naturally leads us to the
consideration of a non-perturbative power correction, \be
\Delta(c)=G(c)K a_0^{-b}e^{-\f{2}{a_0}}+ \,\,\ldots\,.\ee Note
that, at leading order, one can fix the $c$ dependence of the
'condensate' $G(c)$, such that the total result
$\mathcal{R}(c)+{\Delta}(c)$ is independent of $c$, as it should
be.

It remains to be seen, to what extent our results will change if
we consider the renormalization scale variation for real QCD (with
nonzero $\b{1},\b{2},\ldots$).


\begin{thebibliography}{1}

\bibitem{th77}
G't~Hooft.
\newblock Lectures given at Int. School of Subnuclear Physics, Erice, Sicily,
  Jul 23 - Aug 10, 1977.


\bibitem{w69}
K.~G. Wilson.
\newblock {\em Phys. Rev.}, 179:1499--1512, 1969.

\bibitem{svz79}
M.~A. Shifman, A.~I. Vainshtein, and V.~I. Zakharov.
\newblock {\em Nucl. Phys.}, B147:385--447, 1979.

\bibitem{m85}
A.~H. Mueller.
\newblock {\em Nucl. Phys.}, B250:327, 1985.

\bibitem{b98}
M.~Beneke.
\newblock {\em Phys. Rept.}, 317:1--142, 1999.

\bibitem{bz92}
M.~Beneke and V.~I. Zakharov.
\newblock {\em Phys. Rev. Lett.}, 69:2472--2474, 1992.

\bibitem{p78}
G.~Parisi.
\newblock {\em Phys. Lett.}, B76:65--66, 1978.

\bibitem{aw95}
G.B.~Arfken and H.~J. Weber.
\newblock {\em Academic Press}, Mathematical Methods For
Physicists.

\bibitem{s84}
P.~M. Stevenson.
\newblock {\em Nucl. Phys.}, B231:65, 1984.

\bibitem{b93}
D.~J. Broadhurst.
\newblock {\em Z. Phys.}, C58:339--346, 1993.

\end{thebibliography}
\end{document}